\documentclass{ws-ijmpa}
\usepackage[super,compress]{cite}
\usepackage{graphicx}
\input paperdef
\newcommand{\mr}{\mathrm}
\newcommand{\EPJ}{{\it Eur. Phys. J. }}
\newcommand{\AIP}{{\it AIP Conf. Proc. }}
\newcommand{\JHEP}{{\it JHEP }}
\newcommand{\PR}{{\it Phys. Rev. }}
\newcommand{\PL}{{\it Phys. Lett. }}
\newcommand{\PRL}{{\it Phys. Rev. Lett. }}
\newcommand{\CPC}{{\it Comput. Phys. Commun. }}
\newcommand{\low}{$10^{33}\,$cm$^{-2}\,$s$^{-1}$}
\newcommand{\high}{$10^{34}\,$cm$^{-2}\,$s$^{-1}$}

\begin{document}
{}
%
\catchline{}{}{}{}{}
%

\title{\bf REVIEW OF CENTRAL EXCLUSIVE PRODUCTION OF THE HIGGS BOSON BEYOND THE 
STANDARD MODEL}

\author{MAREK TA\v{S}EVSK\'{Y}}

\address{\it Institute of Physics of the Academy of Sciences of the Czech 
republic, \\
Na Slovance 2, 18221 Prague, Czech republic\\
Marek.Tasevsky@cern.ch}

\maketitle


\newcommand{\sixooo}{600 \ifb}
\newcommand{\sixoooeff}{600 \ifb\,eff$\times2$}

\begin{abstract}
We review activities in the field of theoretical, phenomenological and 
experimental studies related to the production of the Higgs boson in central 
exclusive processes at LHC in models beyond Standard Model. Prospects in the 
context of the Higgs boson discovery at LHC in 2012 and of proposals to build
forward proton detectors at ATLAS and CMS side are summarized.

\end{abstract}

\ccode{PACS numbers:
      {13.85.-t}{ high-energy reactions, hadrons-induced}   \and
      {14.80.Bn}{ Higgs bosons, Standard Model} \and
      {14.80.Da}{ Supersymmetric Higgs bosons}}

\section{\bf Introduction}

The central exclusive production (CEP) of the Higgs boson is an exciting area 
of research. It received a great deal of attention over the last two decades 
from theorists as well as experimentalists, see Refs.~\citen{KMR0,ar,KMRProsp,acf,DKMOR,KMRbsm,HarlandLang:2013jf}. The exclusivity and hence a relatively
simple experimental signature makes this process unique,
but the excitement perhaps arises from a question that everybody starting
to explore this process has to ask: what does the exclusivity actually cost? On 
theory side the price is a rather low cross section. On experiment side we pay
by rather low signal selection efficiencies. But the balance is put back by a
favorable signal-to-background ratio. The Higgs boson with mass close to 
125.5~GeV has been discovered at the LHC in 2012 by the ATLAS \cite{HdiscA} and CMS 
\cite{HdiscC} experiments. Preliminary studies of its properties (see for 
example a global analysis collecting all available LHC data in 
Ref.~\refcite{HiggsSignals}) suggest the Higgs boson is compatible with the 
Standard Model (SM), nevertheless there is still room for models of New Physics.
In this review aspects of the search for the Higgs boson produced exclusively 
at the LHC in particular in models beyond the Standard Model (BSM) are 
summarized. 

Theoretically this process has been discussed since the early nineties.
The interest was renewed by the KMR group \cite{KMR0} which provided the 
first realistic evaluation of the expected cross section for the SM Higgs boson.
For several years this process formed the core of the physics
case for the FP420 project \cite{FP420} and its offsprings AFP \cite{AFP} (ATLAS
Forward Proton) in ATLAS and CT-PPS \cite{PPS} (CMS-Totem 
Precision Proton Spectrometer) in the CMS and Totem experiments. For various
reasons the main focus of the FP420 project was installing forward proton 
detectors at 420~m from the interaction point (IP) from ATLAS and/or CMS, while 
the AFP and CT-PPS projects presently concentrate on installing forward proton 
detectors around 220~m. A combination of both locations is viewed as ideal, and many analyses therefore use the combined
acceptance as a basis for calculations and estimates. If forward proton 
detectors (FPD) are installed symmetrically on the left (L) and right (R) side 
of the IP, the number of possible configurations amounts up to
four: two symmetric ones, 420(L)+420(R) and 220(L)+220(R), and two asymmetric 
ones, 420(L)+220(R) and 220(L)+420(R). 
Although the exact positions of the AFP and CT-PPS are 210~m and 240~m, 
respectively, we will denote them 220 in the following. Because there are
three possibilities of installation of the FPDs (stations 
at 220~m only, at 420~m only, and at 220~m and 420~m), the individual four 
configurations form three different acceptance scenarios (and hence different 
kinematic conditions) which will be denoted as ``220" for 220(L)+220(R), 
``420" for 420(L)+420(R) and ``total" for the sum of all four configurations.

\section{\bf Status of Proposals of Forward Proton Detectors at LHC}

The physics, experimental, hardware and integration aspects studied during the 
R\&D phase of the FP420 project were comprehensively summarized in 
Ref.~\refcite{FP420}. The FP420 was a collaboration of ATLAS, CMS and other 
groups which split into two independent projects, AFP in ATLAS and CT-PPS in
CMS and Totem, which concentrate on installing stations only at around 220~m 
from the interaction point, IP, while considering stations at 420~m as a 
possible upgrade. 

Designs of both the AFP and CT-PPS projects are similar. They will use 
3D-Silicon tracking detectors to precisely measure the four-momentum loss of 
the diffractively scattered proton, $\xi$, and either quartz bars, gas detector 
or diamond sensors to measure the time-of-flight (ToF) of the deflected protons 
between the interaction point and the timing detectors. The ToF determines the 
interaction point in the beam direction if (and only if) the protons detected
at opposite sides of the IP came from the same interaction, thus reducing 
pile-up background. Both AFP and CT-PPS (Stage I) house the detectors in Roman 
pots as a means of moving them very close to the beams, as in Totem experiment 
and ALFA (FPD in the ATLAS detector).
At a later stage CT-PPS plans to use a sideways-moving beam pipe as
used at PETRA collider at DESY (Hamburg).
The CT-PPS project has been
endorsed by the LHC Committee in September 2013; the AFP project is close to an 
approval by the ATLAS collaboration by the end of 2014.

\section{\bf Central Exclusive Production of Higgs boson in SM and MSSM}

A comprehensive review and summary of the central exclusive production of 
various processes has been released recently in Ref.~\refcite{CEPreview}.
Here we outline only the main aspects of exclusive production. 

The process is defined as $pp\rightarrow p\oplus\Phi\oplus p$
where all of the energy lost by the protons during the interaction
(a few per cent) goes into the production of the central system, $\Phi$. The 
final state therefore consists of a centrally produced system (e.g. dijet, 
heavy particle or Higgs boson) coming from a hard subprocess, two very forward 
protons and no other activity. The '$\oplus$' sign denotes the regions devoid 
of activity, often called rapidity gaps. A simultaneous detection of both 
forward protons and the central system opens up a window to a rich physics 
program covering not only exclusive but also a variety of QCD, Electroweak and
BSM processes (see e.g. Refs.~\citen{KMRProsp,CMS-Totem,FP420,
AFP,PPS,CEPreview,anomalous,kp1,bcfp,mt1,technipions}). Such measurements can
put constraints on the Higgs sector of Minimal Supersymmetric SM (MSSM) and 
other popular BSM scenarios \cite{KKMRext,diffH1,diffH2,diffH3,CLP,fghpp,ismd,triplet,eds09,dis11,dis12,je2,CEPW,HaoSun,Inan,Senol}. 
In the SM the CEP of Higgs 
bosons has been studied in 
Refs.~\citen{HarlandLang:2013jf,CMS-Totem,CLP,diffH1,cr1,ismd05,ATLASnote,smhww}. 

CEP is especially attractive for three reasons: firstly, if the 
outgoing protons remain intact and scatter through small angles then, to a very 
good approximation, the primary di-gluon system obeys a $J_z=0$, $\cC$-even, 
$\cP$-even selection rule~\cite{KMRmm,KKMR}. Here $J_z$ is the projection of 
the total angular momentum along the proton beam axis. This therefore allows
a clean determination of the quantum numbers of any observed resonance. Thus, 
in principle, only a few such events are necessary to determine the quantum 
numbers, since the mere observation of the process establishes that the 
exchanged object is in the  $0^{++}$ state. Secondly, from
precise measurements of the proton momentum losses, $\xi_1$ and
$\xi_2$, and from the fact that the process is exclusive, the mass
of the central system can be measured much more precisely than from
the central detector, by the so-called
missing mass method \cite{ar}, $M^2=\xi_1\xi_2 s$ ($s$ is the square of the proton-proton 
center-of-mass energy) which is independent of the
decay mode. Thirdly, in CEP, particularly for the so-far elusive $b\bar b$ mode, the 
signal-to-background (S/B) ratios turn out
to be close to unity, if the contribution from pile-up is not considered. 
This advantageous S/B ratio is due to the
combination of the $J_z=0$ selection rule, the potentially excellent
mass resolution, and the simplicity of the event signature in the
central detector. Another important feature of forward proton
tagging is the fact that it enables the strongest decay modes, namely $b\bar b$,
$WW^{(*)}$, $ZZ^{(*)}$ and $\tau\tau$ to be observed in one process. In this 
way, it may be possible to access the Higgs boson coupling to bottom quarks. 
This may be challenging in conventional search channels at LHC due to large QCD 
backgrounds, even though $\hbb$ is the dominant decay mode for a light 
SM Higgs boson. Here it should be kept in
mind that access to the bottom Yukawa coupling will be crucial as an
input also for the determination of 
Higgs couplings to other particles~\cite{HcoupLHCSM,HcoupLHC120,LHCHiggsX1,LHCHiggsX2}.

The last few years witnessed a fair development in the calculations of 
the CEP cross sections concerning both the hard matrix element (see 
Refs.~\citen{acf,mrw,cf,KMRIns,KRS1,Harland-Lang:2013xba,Pasechnik,Ryutin,R+P,Cudell}) and the 
so-called
soft absorptive corrections and soft-hard factorization breaking effects 
(see Refs.~\citen{kmrf,nns2} for details and references). There are basically 
four groups providing calculations of the cross-sections for the CEP of the SM Higgs
boson, whose most recent predictions are in Refs.~\citen{HarlandLang:2013jf,Pasechnik,Ryutin,R+P,Cudell}. 
Their approaches differ in a number of aspects (e.g. 
different soft survival probabilities, different scales in the Sudakov 
derivative or even its complete absence). Predictions differ within the same 
group, depending on input parameters, e.g. parton density functions (PDF), but 
if taken globally, predictions of all four groups range between roughly 0.5~fb 
and 2~fb, leading to an uncertainty factor of three which is much progress
compared to a situation in the previous decade. 
Since the KMR calculations i) successfully describe existing 
data, ii) are maintained and improved and iii) form a basis of all 
analyses reviewed in this document, we briefly summarise the recent development,
while a complete review can be found in Ref.~\refcite{CEPreview}. 
Calculations of the combined enhanced- and eikonal-soft survival
factor give lower values than the value 0.03 used in several analyses, e.g.
in Refs.~\citen{CLP,diffH1,diffH2,diffH3,ATLASnote}. Also taking a more
appropriate factorization scale $M$ (rather than $\approx 0.62 M$) in 
calculating Sudakov suppression almost halves the CEP cross section (of both
the signal and the background) \cite{cf}. 
On the other hand as discussed in \cite{KMRH,HarlandLang:2010ep} we may expect
the cross section to be increased by higher order corrections and by using the
CTEQ6L \cite{CTEQ6L} Leading Order (LO) proton PDF that give the best agreement of the CEP
calculations with CDF data on exclusive $\gamma\gamma$ production 
\cite{CDFgg}. The combined effect of all changes is estimated to be rather small.
Ways to test the theoretical formalism at LHC, 
with or without forward proton tagging, are summarized in Ref.~\refcite{early}.

\subsection{$\Hbb$}\label{Summ}
There are three independent analyses which have been studying the feasibility 
and significance of the CEP Higgs boson decaying into $\bb$ in the SM. They  
have all arrived at an almost identical set of cuts to select the signal and 
suppress all types of background. In Ref.~\refcite{CMS-Totem} experimental 
efficiencies have been obtained in the range of masses between 100 and 300~GeV 
using a fast CMS simulation, in Ref.~\refcite{ATLASnote} results for one mass 
point of 120~GeV were based on a fast ATLAS simulation and in Ref.~\refcite{CLP}
one mass point at 120~GeV has been studied using generator-level quantities that
have been smeared using resolutions published in the ATLAS TDR~\cite{ATLASTDR}.
After applying all these cuts, the numbers of events for the signal, for the 
irreducible backgrounds (defined as coming from the CEP and Double Pomeron 
Exchange (DPE) processes) and 
for the overlap backgrounds (defined later) obtained at the mass of 120~GeV are
very similar among the three analyses, with the exception of 
Ref.~\refcite{CMS-Totem} where the overlap background numbers are higher due to
not applying a cut on the number of charged tracks outside the dijet 
system (see later).
The numbers of CEP signal and background events are estimated by ExHuMe MC 
event generator \cite{ExHuMe}, the DPE background by Pomwig \cite{Pomwig} and 
the overlap backgrounds by Pythia \cite{Pythia} or Herwig \cite{Herwig}. For an 
integrated luminosity of 30 \ifb and for the FPDs at 420~m and 
total acceptance scenarios separately, they amount to 2 and 3 for the signal, 
3 and 4 for the CEP backgrounds and are negligible for the DPE backgrounds. The
remaining overlap backgrounds depend non-linearly on the instantaneous 
luminosity and they were estimated to be negligible at luminosity of \low, 
while they amount to roughly 20 and 40 events at 
luminosity of \high for the two FPD acceptance scenarios. 
Significances coming from these signal and 
background numbers are moderate and they would have to be improved to
attain more favorable prospects. Ways to improve these significances are 
discussed in Refs.~\citen{diffH3,CLP}. For example we can surely expect 
improvements in the gluon-jet/$b$-jet mis-identification probability $P_{g/b}$.
In the original analyses in Refs.~\citen{CLP,diffH1,diffH2,diffH3,ATLASnote} 
a conservative approach has
been followed by taking the maximum of two values available at that time in 
ATLAS and CMS, which was $P_{g/b}=$1.3\% used in ATLAS. Meanwhile new 
developments were reported in reducing the light-quark-b 
mis-identification probabilities in ATLAS \cite{ATLAS-bjet} and CMS \cite{CMS-bjet}. Other possibilities 
to improve the significances in searching
for the SM Higgs in CEP are a possible sub-10 ps resolution 
for timing detectors, the use of multivariate techniques or a further 
fine-tuning or optimization of the signal selection and background rejection 
cuts, thanks to the fact that the mass of the SM-like Higgs boson is already 
known with a relatively high precision. The known Higgs boson mass can also 
greatly facilitate proposals for a dedicated L1 trigger to efficiently save 
events with the CEP $\Hbb$ candidates. Proposals made in 
Ref.~\refcite{Hbb-triggers}, well before the SM-like Higgs boson discovery, can 
thus be further optimized.

\subsubsection{Pile-up effects}\label{PU}
At luminosities greater than \low, high energy interactions
are accompanied by a number of soft interactions in the same bunch-crossing, so called 
pile-up events. The most dangerous combination arises from an 
overlap of a non-diffractive (ND) event with a hard scale (e.g. a dijet event 
with jet transverse momentum as the hard scale in the case we are searching for 
the $\Hbb$ signal) with two additional Single Diffractive (SD) events each 
having a leading proton inside the acceptance of FPDs. The overlap of these 
three events can resemble a signal event and is the most prominent source of 
background in this channel at high instantaneous luminosity (around \high). At 
low instantaneous luminosity (around \low), the overlap background is at a 
per-mil level and the CEP backgrounds dominate (numbers are given in the previous 
subsection). 

In all three analyses in Refs.~\citen{CMS-Totem,ATLASnote,CLP} the quantitative 
effect of
overlaid pile-up events was assessed for the signal sample, for the diffractive 
background sources and for the ND $\bb$ dijet background. It was studied by
mixing, event-by-event, a certain number of minimum bias events with
one signal or one background event. This number is determined from Poisson 
statistics; it depends linearly on luminosity and is model-dependent since the
total $pp$ cross section at $\sqrt s =14$~TeV has not yet been measured. 
The effect of the overlap background is quantified by a
process-independent probability of a fake double-proton tag, i.e. of
detecting two protons in FPDs, each on opposite sides of the IP,
caused by outgoing protons from pile-up events. The results are
summarized in Refs.~\citen{ATLASnote,CLP}, here only the main facts are 
summarized. 

At a luminosity of \low, where the average
number of pile-up events per bunch crossing is 3.5 (including elastic events), 
the probability of an event to have a fake double-proton tag caused by pile-up
protons is a few per-mil for all three FPD configurations. At 
$\cal L \sim$~\high, where the 
average number of pile-up events per bunch crossing is 35 (including elastics), 
the probabilities reach 7\%, 21\% and 23\% for the 420$+$420, 220$+$220 and 
420$+$220 configurations, respectively, for assumed distances from the 
beam center of 2.5~mm at 220~m, and 4~mm at 420~m including a dead zone of 1~mm.

The numbers of fake proton pairs per bunch crossing, N/BX, as a function of the 
number of pile-up events per bunch crossing, $N^{\mr P\mr U}$, is purely
combinatorial and well described by the following formula 
\cite{ATLASnote}:
\begin{equation}
N/BX = 2e^{-\mu_{\mr S\mr S}}(\cosh(\mu_{\mr S\mr S})-1) + 1-e^{-\mu_{\mr D\mr S}}
\label{fakerates}
\end{equation}
Here $\mu_{\mr S\mr S}=A_{\mr S\mr S}*N^{\mr P\mr U}$ and 
$\mu_{\mr D\mr S}=A_{\mr D\mr S}*N^{\mr P\mr U}$ 
denote the probability to see pile-up protons in the FPDs on one
side only (``SS'', single-sided) and on both sides (``DS'', double-sided), 
respectively. 
As is evident from the formula, the number N/BX crucially depends on the 
probability $A_{\mr S\mr S}$. There are several estimates of this probability for
FPDs at both, 420~m and 220~m available in literature which can be compared.  
In Ref.~\refcite{ATLASnote} a high statistics Pythia sample of minimum bias 
events and 2-dimensional ($\xi$, $p_T$)-FPD acceptances gave 0.9\% and 1.7\% 
for $A_{\mr S\mr S}$ at 420~m and 220~m, respectively, and negligible numbers for 
$A_{\mr D\mr S}$ at both, 420~m and 220~m. The latter fact is due to the absence 
of the DPE processes in Pythia. In Ref.~\refcite{CLP} only basic cuts 
0.005$<\xi_1<$0.018, 0.004$<\xi_2<$0.014 and 0.02$<\xi<$0.2 were used as rough 
estimates of the 420 and 
220 acceptances, respectively, and for these, 1.1\% and 2.4\% were obtained with
PYTHIA~6.4, while 1.2\% and 3.1\% were reached by PHOJET~1.2. In 
Ref.~\refcite{CMS-Totem} where the distances of the active region of the 
detectors were 1.5~mm and 4.5~mm for the 220 and 420 stations, respectively, 
and 2-dimensional ($\xi$, $p_T$)-FPD acceptances and PHOJET~1.2 were used, the 
probabilities of 1.0\% and 3.1\% were reported. Finally, a theoretical estimate
was made based on a model with a complete set of multi-Pomeron
vertices. This model is described in Ref.~\refcite{KMR-multicompmodel},
where it is shown to be consistent with the HERA data on the
triple-Pomeron vertex, and the estimates were based on the same basic
$\xi$ cuts as used in Ref.~\refcite{CLP} are 1.0\% and 2.5\%.
These numbers critically depend on the normalization of the leading
proton $\xi$ spectrum, which has not yet been measured at $\sqrt{s} = 14$~TeV 
for which the above estimates are given.

At $\sqrt{s} = 14$~TeV, cross-sections of soft SD events or ND dijet events with
$\hat{p_\mr T} > 30$~GeV range around 10~mb or tens of $\mu$b, 
respectively, so reducing such huge backgrounds down to a tolerable level,
comparable or smaller than the cross section of the signal after applying all
the stringent cuts, is a real challenge. 

All signal selection and background rejection cuts are described in 
Refs.~\citen{CMS-Totem,ATLASnote,CLP}. As an example, in Fig.~\ref{PUcuts} we
show two quantities that provide an excellent separation between the signal and
the overlap background and are thus used to significantly suppress the overlap 
background.  

\begin{figure}[h]
\includegraphics[width=6.3cm]{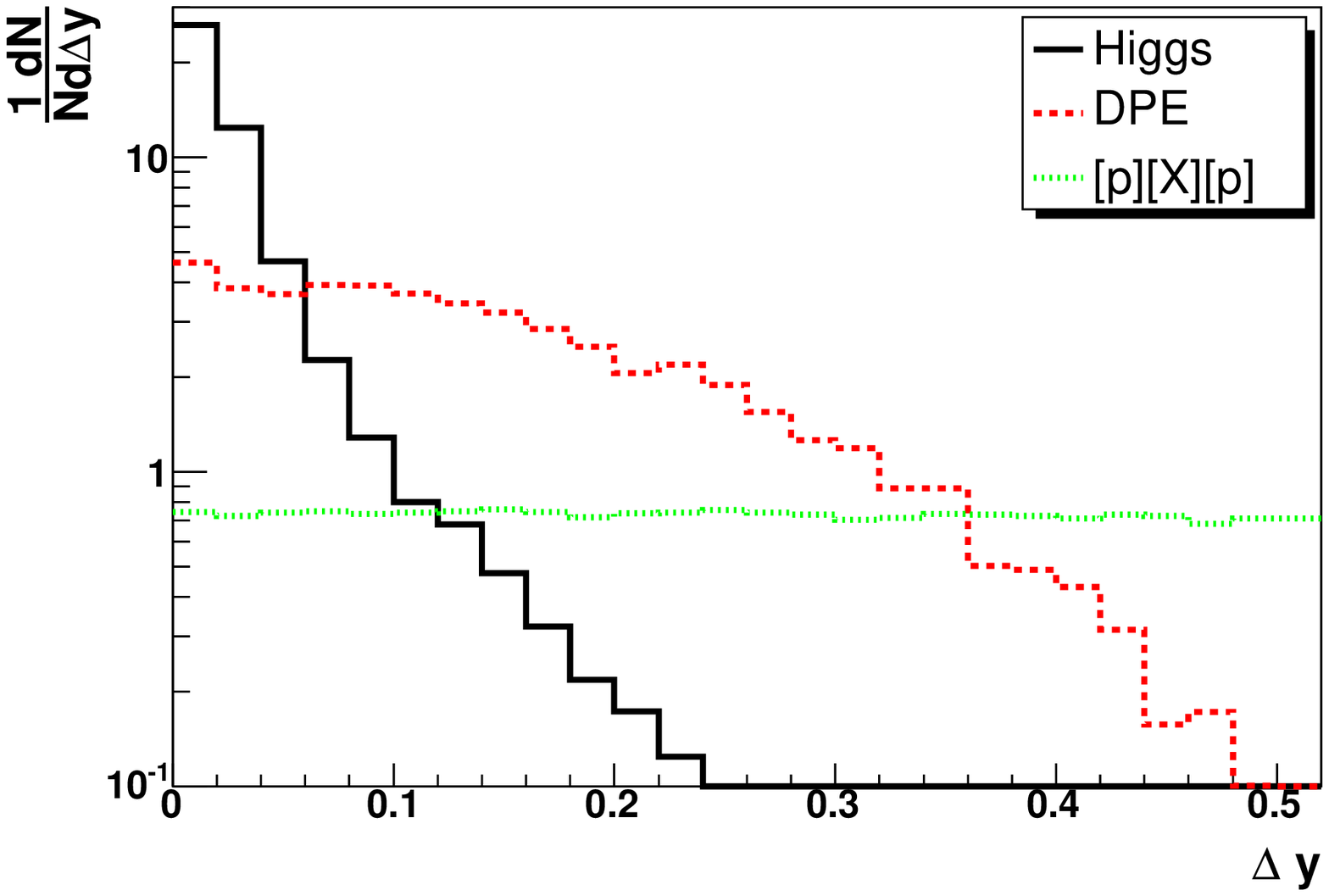}
\includegraphics[width=6.3cm]{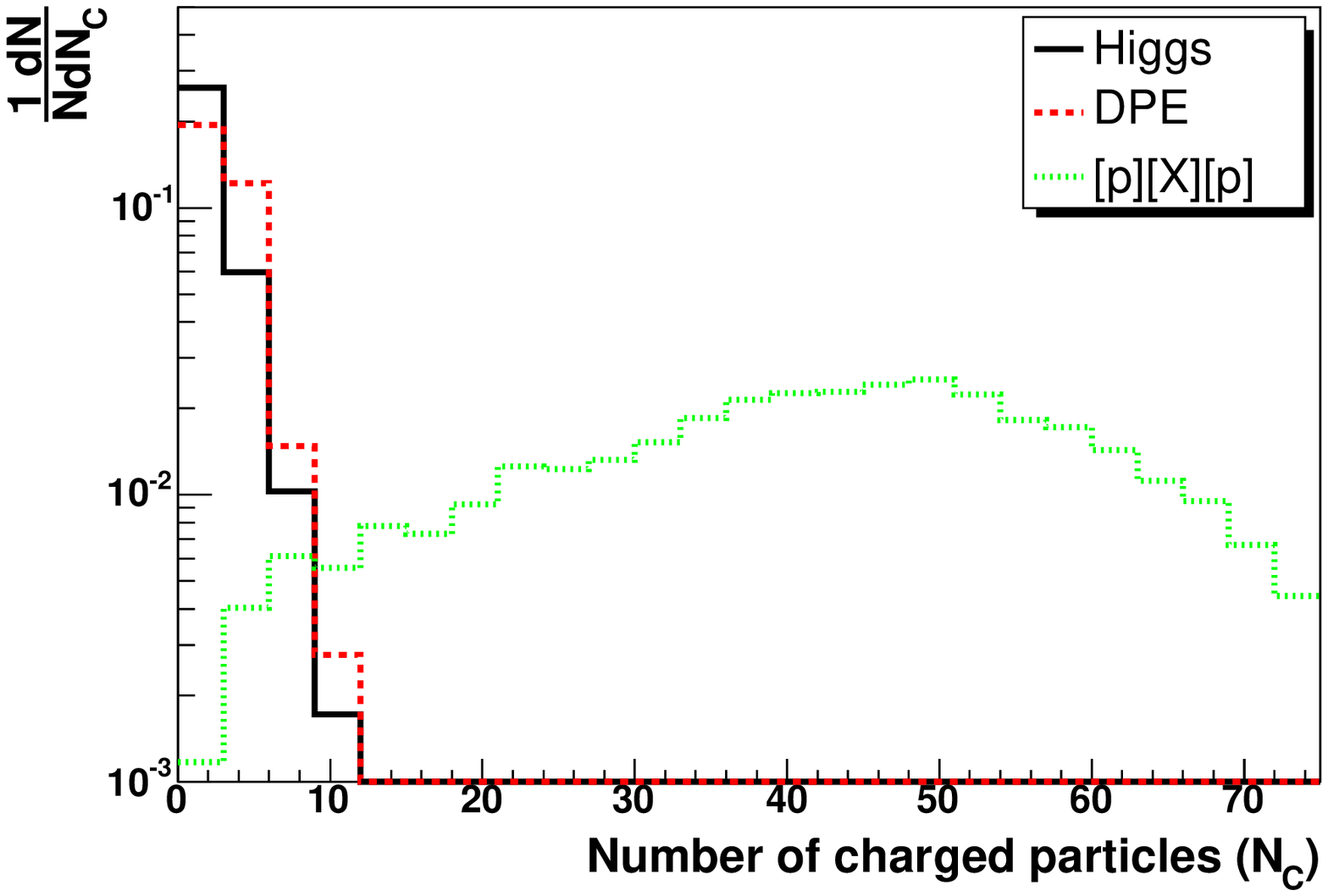}
\caption{The $|\Delta y|$ (left) and N$_{\mr c}$ (right) distributions are shown
for the signal, DPE and overlap backgrounds. Events were generated by the
ExHuMe MC event generator with momenta of particles smeared with ATLAS detector 
resolutions, and use the cone jet algorithm with a cone radius of 0.7. Taken
from Ref.~22.}
\label{PUcuts}
\end{figure}

The $|\Delta y|$ variable is defined as a difference between pseudorapidity
of the central system calculated from the central detector and from the FPD,
the latter based on the precise measurements of $\xi$. The  N$_{\mr c}$ variable 
is a number of charged particles outside the dijet system and thus gives an 
estimate of the number of charged particles not associated with the hard-scale 
process. The multiplicity difference for the signal and overlap background can 
then be accounted for by two facts: firstly the central system in 
non-diffractive 
events is color-connected with proton remnants, therefore the total number of 
produced particles is higher than that in a diffractive event. 
Secondly, basically for the same reason, the underlying event producing soft 
particles may be present in the non-diffractive event, while it is believed to
be significantly suppressed in the diffractive event.

Another very powerful means to suppress the overlap background is exploiting
fast timing detectors whose goal is to distinguish whether the proton detected 
in a FPD originates from a hard-scale event or from a pile-up event. The ToF 
difference between the two protons gives the $z$-position of the collision, to 
match with the central event. Fast timing detectors with an expected resolution 
of around 10~ps and hence $z$-resolution $\sim$2~mm are 
part of the AFP and CT-PPS projects. Monte Carlo studies \cite{ATLASnote,CLP} 
suggest that with nominal running conditions a fair rejection is possible of 
events with fake protons. If a 2$\sigma_t$ cut ($\sigma_t$ is resolution of
the ToF detector) is applied on the time 
measurement, the rejection is 19 (18, 16 and 15) at an instantaneous luminosity
of 1 (2, 5 and 10) $\times$~\low, respectively.

\subsubsection{BSM scenarios}
Numerous models of new physics require an extended Higgs sector. A frequently 
used extension of the SM is the MSSM~\cite{susy1,susy2,susy3}, where the Higgs 
sector consists of five physical states (two Higgs doublets are required). At 
the lowest order the MSSM Higgs sector is $\cp$-conserving, containing two 
$\cp$-even bosons, the lighter $h$ and the heavier $H$, a $\cp$-odd boson, $A$, 
and the charged bosons $H^\pm$. It can be specified in terms of the gauge 
couplings, the ratio of the two vacuum expectation values, 
$\tb \equiv v_2/v_1$, and the mass of the $A$ boson, $\MA$. The Higgs sector of 
the MSSM is affected by large higher-order corrections (see for example 
Refs.~\citen{reviews1,reviews2,reviews3} for reviews), which have to be taken 
into account for reliable phenomenological predictions. 

In the last decade, detailed analyses of the CEP of the Higgs boson in MSSM have
been presented in four basic papers~\citen{CLP,diffH1,diffH2,diffH3}. Results of
Refs.~\citen{diffH1,diffH2,diffH3} are based on the same set of experimental
acceptances and efficiencies, namely that discussed in Ref.~\refcite{CMS-Totem},
and the evolution between the results of Ref.~\refcite{diffH1} and 
Ref.~\refcite{diffH3} is described later. Since results in Ref.~\refcite{CLP}
and Refs.~\citen{diffH1,diffH2,diffH3} are based on 
independent analyses, here 
we briefly outline the main differences in approaches and compare results where 
possible. The differences are as follows: 

\begin{romanlist}
\item {\sc FPD acceptances.} The results in Ref.~\refcite{CLP} are based on the 
ATLAS detector and FPD acceptances at the ATLAS side of the IP, 
while the results in Refs.~\citen{diffH1,diffH2,diffH3} are based on the CMS 
acceptances. A non-negligible difference in the FPD acceptances is observed for
the asymmetric 420+220 configuration where the acceptance at CMS side is lower
by roughly 30\% than that at ATLAS side, if the same distances from the beam 
center are considered. This is caused by the horizontal (rather than vertical)
plane of the crossing angle of the beams at CMS IP.
\item {\sc Decay modes.} In Ref.~\refcite{CLP} only the $b\bar b$ decay mode is 
studied, while in Refs.~\citen{diffH1,diffH2,diffH3} all three main decay modes,
namely $\bb$, $WW$ and $\tau\tau$ are studied. 
\item {\sc L1 triggers.} In Ref.~\refcite{CLP}, three L1 triggers are considered
for the symmetric 420+420 configurations, while for the asymmetric 420+220
configuration, a L1 trigger based on the information from the FPD at 220~m is
used and this trigger is assumed to be fully efficient. The first L1 trigger 
for the symmetric FPD events is a low
$p_T$ muon trigger of 6~GeV in addition to a 40~GeV jet. The second is a 
rapidity gap trigger where gaps would be defined by requiring a veto in very 
forward detectors (such as ZDC and LUCID in ATLAS or ZDC, CASTOR and Totem T1 
and T2 telescopes on the CMS side). This trigger requires low luminosities - 
the trigger efficiency is estimated to be $\sim$ 17\% at \low and $\sim$ 2\% at 
$2\,\times$~\low. 
The third trigger allows a high, fixed L1 
rate for 40~GeV jets which is then substantially reduced at L2 by utilizing 
information from stations at 420~m. The L1 rates of 25~kHz and 40~kHz are 
examined. In Ref.~\refcite{diffH1} 
any of the following four L1 triggers can be used. The first takes the signal
in one of the stations at 220~m and at least two jets with $p_T > 40$~GeV. As 
was shown in Ref.~\refcite{Monika-trigger}, for this trigger a tolerable L1 
bandwidth of 1~kHz may be kept only up to luminosities of about 
${\cal L} \sim 2\,\times$~\low due to the 
overlap background. The second
trigger requires a jet with  $p_T > 40$~GeV and one muon with  $p_T > 3$~GeV,
both measured in the central detector. The third trigger requires at least two
jets with $p_T > 90$~GeV in the central detector, so this trigger is useful
only for high-mass searches. The fourth trigger, requiring electrons or muons 
in the central detector, is focused on retaining mainly events with $W$ bosons
decaying leptonically or semileptonically. While corresponding cuts are used
at analysis level to mimic the triggers, they are assumed to be fully 
efficient.   
\item {\sc Background treatment.} As stated in the previous item, the two 
analyses only overlap in the $b\bar{b}$ decay mode, therefore the background
treatment can only be compared for this decay mode. In Ref.~\refcite{diffH1} 
background sources of the $WW$ and $\tau\tau$ decay modes are also carefully
discussed and these are summarised in separate subsections below. In both 
analyses, the same sources of background are considered for the $\bb$ decay 
mode, namely CEP $gg\rightarrow gg/\bb$, DPE $gg\rightarrow \bb$ and overlap 
backgrounds coming from an overlap of a hard scale event with pile-up events.
\begin{romanlist}
\item {\it Non-overlap backgrounds.}
In Ref.~\refcite{CLP}, the contributions of the above sources of background are
estimated using the LO MC event generators, namely ExHuMe for the exclusive
processes and Pomwig with the H1 diffractive PDFs (so called H1 2006 DPDF Fit B)
for DPE processes. In Refs.~\citen{diffH1,diffH2,diffH3} the background 
contributions 
are calculated analytically for the CEP and DPE processes including 
higher-order contributions and the following four most prolific sources of the 
CEP background have been 
considered in : i) LO $gg\rightarrow gg$ where gluons may be misidentified as 
b-jets. Note that here a conservative value of  $P_{g/b}=$1.3\% has been used 
for this misidentification probability. ii) An admixture of $|J_z|=2$ 
production, arising from non-forward going protons, which contributes to the LO 
$gg\rightarrow \bb$ background. iii) Since the b-quarks have non-zero 
mass there is a contribution to the $J_z=0$ cross section of the order of 
$m_b^2/E_T^2$. iv) Finally the NLO $gg\rightarrow \bb g$ contribution which 
for hard gluon radiation at large angles does not obey the selection rules. The 
latest results \cite{Shuvaev,screen} suppressing the contribution in the LO by 
a factor of roughly two are included in the calculation of the total background 
cross section in the point iv. To suppress the contribution of the DPE 
$gg\rightarrow b\bar{b}$ background, an additional cut on the polar angle of 
$b$-jets in the $\bb$ rest frame, 
$60^{\circ} < \theta < 120^{\circ}$, has been imposed
after which this background has been estimated to be very small, as
confirmed in Ref.~\refcite{CLP} by using the LO Pomwig generator.   
\item {\it Overlap background.} The effect of the overlap background has been 
elaborated in Ref.~\refcite{CLP}, while in Refs.~\citen{diffH1,diffH2,diffH3}
the overlap background is considered to be negligible after applying all 
stringent 
cuts to suppress this background. As summarised above, at luminosities around
\low, the overlap background may be neglected, however,
by neglecting this background at luminosities \high, one 
anticipates a fair improvement in selecting signal as well as in taming
this background. There are definitely ways to improve significances and they 
were discussed above.  
\end{romanlist}
\item {\sc MSSM parameter space.} In Ref.~\refcite{CLP} one example point is
examined, namely the $(\tan\beta, M_A)$ = (40, 120 GeV), while in 
Refs.~\citen{diffH1,diffH2,diffH3} a full scan over the MSSM parameter space
$(\tan\beta, M_A)$ = (2--50, 100-300 GeV) is performed. 
\item {\sc MSSM benchmark scenarios}. In Ref.~\refcite{CLP} the above example
point is studied in the Mhmax scenario, while in Ref.~\refcite{diffH1} two
benchmark scenarios are examined for the $\bb$ decay mode, namely the
Mhmax and no-mixing scenarios. In Ref.~\refcite{diffH2,diffH3} a number of other
scenarios have been investigated and they will be described later. 
We note that in Ref.~\refcite{diffH1} the Mhmax and no-mixing scenarios were 
also used for the $\tau\tau$ decay mode, while for the $WW$ decay mode
the small $\alpha_{\mr eff}$ scenario is used as the one where the enhancement of 
signal cross sections is highest. 
\end{romanlist}

A comparison of the two independent analyses (described in 
Refs.~\refcite{CLP,diffH1}) has been performed in 
Ref.~\refcite{FP420} for the only point in the MSSM parameter space that has 
been studied in Ref.~\refcite{CLP}, namely at $(\tan\beta, M_A)$ = (40, 120 
GeV) and a good agreement has been found at luminosities where pile-up is not 
an issue (i.e. smaller than $2\,\times$~\low). 
A comparison we make here shows a qualitative agreement also at high 
luminosities 
(\high), if similar conditions are used to analyse the
signal and backgrounds. Both analyses use very similar cuts to select the 
signal and to suppress the backgrounds. If also similar FPD acceptances are
used, very similar signal selection efficiencies can be achieved.
In Ref.~\refcite{CLP} FPD acceptances are provided for several options of the
distance of the detector from the beam center, therefore for this comparison 
we have chosen the option which gives the closest FPD acceptances to those
used in Refs.~\citen{diffH1,diffH2,diffH3} (all FPD configurations are used). 
Concerning backgrounds, on the one hand, as explained above, the overlap background 
is not zero even after applying all stringent suppressing cuts. On the other 
hand, the differences in the non-overlap background treatment summarised above 
have not been fully evaluated, nevertheless, rough estimates show that the 
irreducible backgrounds in Refs.~\citen{diffH1,diffH2,diffH3} turn out to be 
higher than in Ref.~\refcite{CLP} because more effects are included than in the 
LO ExhuMe program, so in the end the deficit of remaining background in 
Refs.~\citen{diffH1,diffH2,diffH3} is partly compensated. 
If finally the L1 triggers are assumed to be fully efficient, both analyses
yield  significances well above $5\,\si$ for the studied point 
$(\tan\beta, M_A)$ = (40, 120~GeV) at any instantaneous luminosity, as observed
in Fig.~2 top in Ref.~\refcite{diffH2}, which is to be compared with Figs.~15 
and 17 b) in Ref.~\refcite{CLP}).

The results presented in Refs.~\citen{diffH1,diffH2,diffH3} show an evolution
in time which can briefly be summarised as follows. The results from 
Ref.~\refcite{diffH1} have already been described above. 
In Ref.~\refcite{diffH2,diffH3} the actual version of the code 
{\tt FeynHiggs}~\cite{feynhiggs1,feynhiggs2,mhiggslong,mhiggsAEC,mhcMSSMlong} 
was always employed for the MSSM cross section and decay width calculations and of the 
code {\tt HiggsBounds}~\cite{higgsbounds1,higgsbounds2} for the evaluation of 
the exclusion regions corresponding to searches for MSSM Higgs bosons at LEP, 
Tevatron and LHC. After Ref.~\refcite{diffH2} more accurate calculations were 
taken of the process associated with bottom-mass terms in the Born amplitude 
contributing to the total background of the $\bb$ mode \cite{Shuvaev,screen}. 
Besides the $\Mhmax$ and no-mixing scenarios other scenarios have been 
investigated, namely in Ref.~\refcite{diffH2} those yielding the
correct amount of the cold dark matter abundance, the so-called Cold Dark 
Matter (CDM) scenarios (see Refs.~\citen{EWPOBPOCDM,CDM} for more details), and 
another model beyond the SM, the so-called SM4 model (see e.g. 
Ref.~\refcite{four-gen-and-Higgs}) with a fourth generation of quarks and 
leptons. 
While the SM4 model is practically ruled out \cite{Djouadi-Lenz} by the recent 
LHC measurements of Higgs-mediated cross sections 
\cite{SM4-indirect1,SM4-indirect2} and direct searches 
\cite{SM4-direct1,SM4-direct2}, the CDM scenarios are still viable. 
In Ref.~\refcite{diffH3}, concentrating on the $\bb$ decay mode, seven other new 
benchmark scenarios have been investigated, all compatible with the mass
and production rates of the observed Higgs boson signal at 125.5~GeV and all 
recently proposed in Ref.~\refcite{newscenarios}. The scenario with the best
prospects for observing the CEP $\Hbb$ process in the MSSM is the so-called 
low-MH scenario shown in Fig.~\ref{lowMH}. In the allowed region (the green 
area in Fig.~\ref{lowMH}) which is given by the combined experimental and 
theoretical uncertainties on the Higgs boson mass (see Ref.~\refcite{Hmunc} for 
more details about the theory uncertainties), significances of at least 
$3\,\si$ are obtained for light Higgs boson masses around 80--90~GeV. This is
a mass region which is slightly more challenging than that so far studied around 120~GeV 
(see more discussion in Ref.~\refcite{diffH3}). We note that 
this region of interest may be ruled out by recent ATLAS searches for the light 
charged Higgs boson \cite{ATLASchargedH} which could not be considered in
Ref.~\refcite{diffH3}. However a dedicated analysis of the exclusion bounds in 
this scenario, including the CMS results, still needs to be carried out. 
We conclude that although room for a comparatively low mass (below 200 GeV) BSM
Higgs boson is decreasing, the heavy BSM Higgs bosons cannot be entirely excluded. 

\begin{figure}[h]
\begin{center}
\includegraphics[width=12cm]{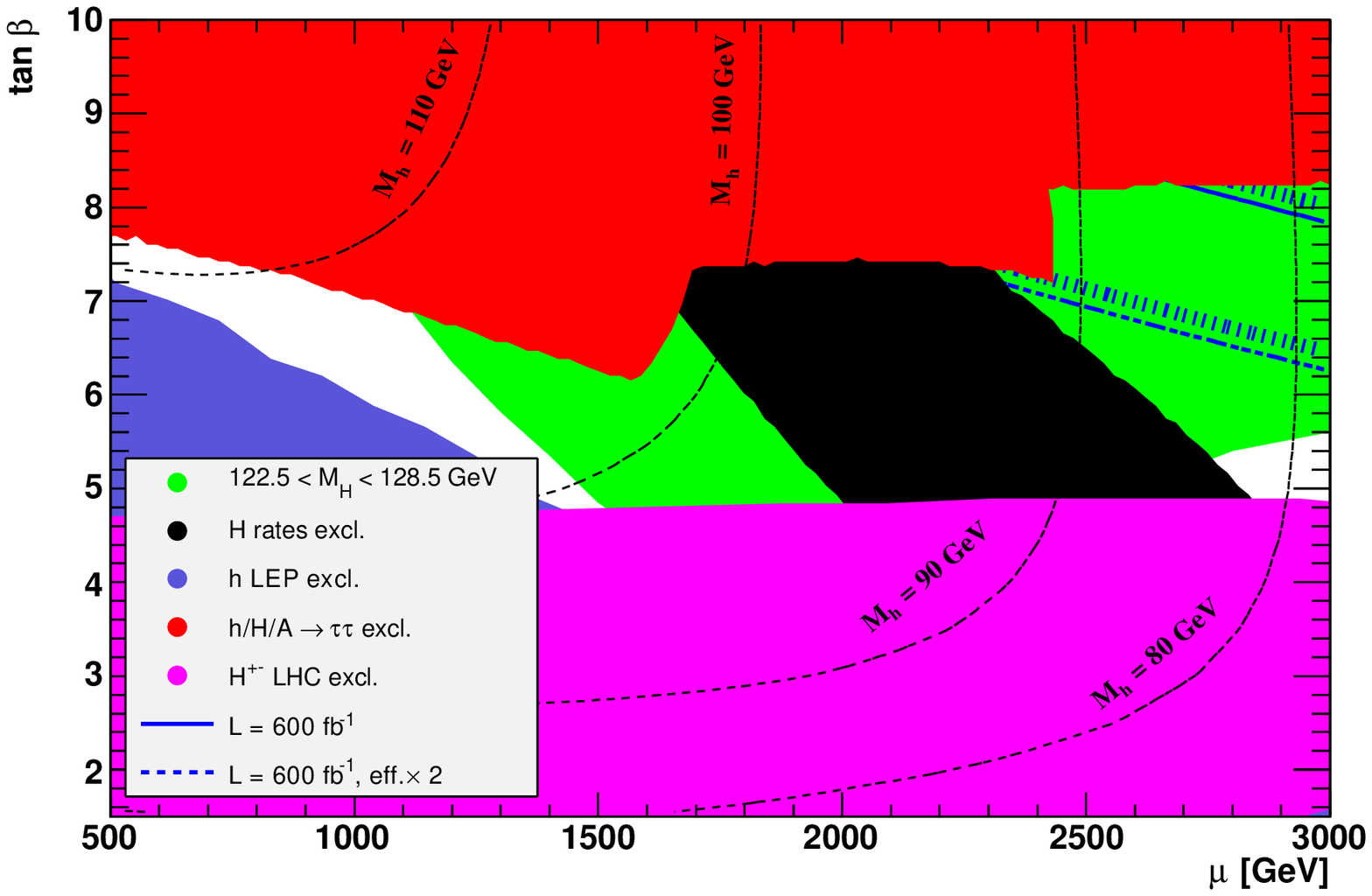}
\caption{Contours of $3\,\si$ statistical significance (solid blue lines) 
for the $\hbb$ channel in CEP at $\sqrt s = 14$~TeV in 
the ($\tb$, $\mu$) plane of the MSSM within the Low-MH benchmark scenario. 
The results are shown for assumed effective
luminosities (see text, combining ATLAS and CMS) of \sixooo\ and \sixoooeff.
The values of the mass of the light $\cp$-even Higgs boson, $\Mh$, are
indicated by dashed (black) contour lines. The dark shaded (blue) region 
corresponds to the parameter region that is excluded by the LEP MSSM Higgs 
searches, the lighter shaded (red), the lighter shaded (pink) and black areas 
are excluded by LHC MSSM Higgs searches in the analyses of 
h/H/A $\rightarrow \tau\tau$, charged Higgs and Higgs rates, respectively. The 
light shaded (green) area corresponds to the allowed mass region 
$122.5 < \MH < 128.5$~GeV. 
}
\label{lowMH}
\end{center}
\end{figure}

Refs.~\citen{diffH2,diffH3} also stress the importance and advantages 
of the CEP process in determining the spin-parity quantum numbers of Higgs 
bosons, and also in searching for a possible $\cp$-violating signal in the Higgs
sector \cite{kmrcp}.


\subsection{\bf \it $H\rightarrow WW^{(*)}$}
This channel does not suffer from any of the difficulties present in the $\bb$ 
channel: suppression of the dominant backgrounds does not rely primarily on 
the mass resolution of the FPDs, background cross sections are calculated with 
a sufficient precision and probably, in the leptonic and semi-leptonic decay 
modes, standard L1 leptonic triggers may be used. 
From experimental point of view, there are three main categories of events
with two W bosons in the final state. 
\begin{enumerate}
\item{\bf Fully leptonic.} Events in which both W bosons decay into either an 
$e$ or a $\mu$ are the simplest and will typically pass the L1 trigger 
thresholds due to the high $p_T$ final state lepton. A portion of the events 
with decays into one or two $\tau$'s will pass the L1 trigger thresholds with 
$\tau$ subsequently decaying leptonically. The inclusive branching ratio (with 
decays either into $e$ or $\mu$ or $\tau$) is 10.3\%. 
\item{\bf Semi-leptonic.} Events in which one W boson decays leptonically and 
the other into two jets can be triggered by the leptonic L1 triggers 
mentioned in the item above. The branching ratio for cases when one W boson 
decays 
into one of the lepton types and the other W boson into two jets is about 
14.5\%, the inclusive branching ratio is therefore about 43.5\%.
\item{\bf Fully hadronic.} The 4-jet decay mode occurs in 46.2\% but it is
unlikely to pass the L1 trigger thresholds without information from the FPDs. 
In addition the QCD background is expected to be overwhelming.  
\end{enumerate}

The backgrounds to the $WW$ decay mode have been thoroughly discussed in 
Ref.~\refcite{diffH1} and also in Refs.~\citen{ATLASnote,smhww,KRS2}. The main
exclusive backgrounds can be divided into two broad groups: i) central 
production of a WW pair from the QED process $\gamma\gamma\rightarrow WW^{(*)}$ 
and ii) W-strahlung process arising from the QCD $gg\rightarrow Wq\bar{q}$ 
subprocess, where the (hadronically) decaying $W^{(*)}$ is faked by the two
quarks. As shown in Ref.~\refcite{ATLASnote,smhww} over a wide range of Higgs
masses the photon-photon backgrounds can be strongly suppressed if the final
state leptons and jets are required to be central and the cut $p_T > 200$ MeV
on the proton transverse momentum measured in the FPDs is
imposed. In general, the gluon-gluon QCD process with the $W$-strahlung is 
expected to be effectively suppressed by requiring the dijet mass to fall in an 
appropriate $W$ mass window, since the background dijet mass distribution is a 
continuum beneath the $W$ mass peak \cite{KRS2}. So for Higgs masses below
the $WW$ threshold (as in the case of the discovered Higgs boson with the measured 
mass of about 125 GeV) where the off-shell $W$-boson is the one decaying 
hadronically, and imposing a $W$ mass window $70<M_W<90$ GeV on the measured 
dijet mass, one should effectively remove all events with hadronically 
decaying $W^{*}$-bosons and enhance the signal-to-background ratio.

There are four analyses in the literature that deal with the $WW^{(*)}$ decay mode.
In Ref.~\refcite{smhww} the signal event yields for the fully leptonic and 
semileptonic channels are presented for the Higgs boson  produced exclusively in
the SM. In Ref.~\refcite{ATLASnote} the event yields are given for the Higgs bosons
decaying semi-leptonically in the $\mu\nu$jj final state and for its most 
important backgrounds, including the overlap source. Both these analyses 
concentrated on the mass point of 160~GeV where the cross-section is largest
for this decay mode in the SM. On the other hand in Ref.~\refcite{diffH1} the 
main emphasis is on investigating prospects of the MSSM, in the vicinity of 
120~GeV, because that is where the enhancement of MSSM 
over SM cross sections is largest (in the small $\alpha_{\mr e\mr f\mr f}$ 
scenario, giving a factor 4). A table of signal selection efficiencies in the range
between 120 and 200~GeV has initially been provided in Ref.~\refcite{CMS-Totem}.
In Refs.~\citen{diffH1,smhww} all important sources 
of irreducible background are discussed and ways to effectively suppress them
experimentally are proposed. All analyses providing signal yields 
(Refs.~\citen{CMS-Totem,smhww,ATLASnote}) are based on the ExHuMe event generator.
With similar FPD acceptances and signal selection cuts, analyses in 
Refs.~\citen{ATLASnote,smhww} arrive at similar signal event yields for a data
taking period corresponding to an integrated luminosity of 30 \ifb. 
According to Ref.~\refcite{smhww} one expects to collect in total 2.3 and one 
Higgs boson decaying semi-leptonically and fully-leptonically, respectively, if
all decay channels are included. 
Slightly less than one Higgs boson candidate with two W-bosons decaying 
semi-leptonically in the $\mu\nu$jj final state is reported in 
Ref.~\refcite{ATLASnote}. This can be viewed as a good agreement given the fact
that all decay modes for each lepton contribute equally. The selection 
efficiencies presented in Ref.~\refcite{CMS-Totem} are twice as high as 
those in Ref.~\refcite{smhww}, due to missing cuts suppressing the exclusive
QED (the cut on $p_T$ of outgoing protons) and QCD (mass window around the W-boson mass) 
backgrounds. Numerical estimates of the backgrounds are given in
Ref.~\refcite{ATLASnote} for the semileptonic channel in the $\mu\nu$jj final 
state. The exclusive QED $\gamma\gamma\rightarrow WW^{(*)}$ turned out to be 
the most important background but can be kept smaller than the signal yield thanks to 
the $p_T$ cut on the proton measurements in FPDs. The overlap background turned out 
to be fully tamable after making use of the exclusivity cuts. These conclusions
are valid for any instantaneous luminosity up to \high.

\subsection{\bf \it $H\rightarrow \tau\tau$}
The Higgs boson branching ratio into two $\tau$'s in SM at a mass of 120~GeV 
being ten times smaller than that into $\bb$ makes this decay mode less
attractive than the other two discussed above. Similarly to the
$WW$ decay mode, there are three classes of events with respect to the decay 
mode of $\tau$, namely the fully leptonic, semi-leptonic and fully hadronic, 
with a significant difference being the higher number of neutrinos in the final 
state. Important experimental advantages compared to the $\bb$ decay mode are 
i) the possibility of using high $p_T$ lepton triggers and ii) that hadronic 
$\tau$-jets have very low particle multiplicities, and are almost exactly 
back-to-back in azimuthal angle. The latter promises to enable us to 
effectively suppress the overlap background even for the hadronic decays. 

Backgrounds are discussed in Ref.~\refcite{diffH1}. The dominant sources
come from two processes: the QED $\gamma\gamma\rightarrow\tau\tau$ which
can be safely neglected after applying cuts on the transverse momenta of
the diffractive protons measured in the FPDs. The cut $p_T > 200$
MeV suppresses the QED background by a factor of 70, while the signal is reduced
by 40\%. The other important background comes from the CEP $gg\rightarrow gg$ 
where the outgoing gluons may mimic the tau. Imposing the same cut as in the
$b\bar{b}$ case, $60^{\circ} < \theta < 120^{\circ}$, and assuming that the 
probability of misidentification of gluons to be tau, $P_{g/\tau}$ is smaller 
than 1/50, this source of background can be safely neglected. Note that in the 
inclusive events, thus in a less clean environment than in the exclusive case 
we are discussing here, $P_{g/\tau}$ has been evaluated to be 1/500 
\cite{g-tau1,g-tau2} and so our assumption is realistic. 

Estimates of signal and background event yields for this decay mode have never 
been published. In Ref.~\refcite{diffH1,diffH2} the same signal selection 
efficiencies as for the $\bb$ decay mode are assumed, while the exclusive QED
as well as QCD backgrounds have been properly calculated. Unpublished results 
\cite{Vlasta} based on exclusivity cuts and including those suppressing both 
exclusive backgrounds discussed above confirm the similarity of the signal 
selection efficiencies between the $\tau\tau$ and $\bb$ modes. Prospects for 
this channel in MSSM have been presented in
Refs.~\citen{diffH1,diffH2} in the same benchmark scenarios as for the $\bb$ 
channel. Although the outlook for taming the overlap backgrounds is more favorable
than for the $\bb$ decay mode, the significances are generally always 
lower.

\section{\bf Central Exclusive Higgs production in NMSSM}

The Next-to-Minimal Supersymmetric Standard Model (NMSSM) extends the MSSM by
introducing a singlet superfield. The Higgs sector in the NMSSM consists of 
three $\cp$-even and two $\cp$-odd neutral Higgs bosons and a charged Higgs 
boson. In this model the $\mu$ problem \cite{muproblem} can be solved and scale 
parameters do not need to be finely tuned as is the case in MSSM. However this 
is only possible if the lightest $\cp$-even Higgs boson $h$ has a mass around 
100~GeV and has SM couplings to gauge bosons and fermions. Consistency with LEP 
limits then results in $h$ primarily decaying via $h\rightarrow aa\rightarrow 
\tau^+\tau^-\tau^+\tau^-$ where $a$ is the lighter of the two pseudoscalar Higgs
bosons present in the NMSSM. Both of these Higgs bosons might have escaped the LEP
searches and they are almost impossible to detect at the LHC by standard tools.
The troublesome low mass region may be covered by CEP. In Ref.~\refcite{nolose}
the above scenario is studied in detail with a scalar Higgs boson of mass 
92.9~GeV decaying into two pseudo-scalar Higgs bosons $a$ with masses of 9.73~GeV.
The $h\rightarrow aa$ decay occurs in 92\% cases and each $a$ then decays
into $\tau^+\tau^-$ with 81\% probability. The analysis procedure is very 
similar to that used in studying the MSSM CEP $\Hbb$ process in 
Ref.~\refcite{CLP}. The signal has been incorporated into the ExHuMe event 
generator, which also served to estimate the effect of irreducible backgrounds 
coming from the CEP $gg\rightarrow gg/\bb$ processes, and POMWIG was used to
estimate the DPE dijet background. The overlap background 
coming from pile-up events is treated as described in subsection~\ref{PU}. 
Finally the exclusive QED backgrounds $pp\rightarrow \tau^+\tau^-l^+l^-$ (where 
$l$ is any charged lepton) were also considered and estimated using MADGRAPH 
\cite{madgraph}. The L1 trigger considered is a single muon with $p_T > 6$~GeV 
(which may have to be increased to 20~GeV at high luminosity). 

After all cuts and using only the 420 FPD configuration (since the stations at 
220~m do not contribute due to the low mass region of interest), effective 
cross sections (after applying all cuts) are 0.07~fb for signal and 0.002~fb 
for the non-overlap background, where the largest source is the DPE dijet 
production, while the QED source is entirely negligible. The overlap background 
is 0.005~fb at luminosities of \high and 100 times lower at \low. These values 
correspond to a central mass window
of 70--110~GeV assuming that such a wide interval will be needed when
operating in a search mode. Once the signal is detected, a much narrower mass
window can be applied thereby reducing the overlap background significantly. 
The small size of this background (compared e.g. to the CEP $\Hbb$ channel where
it represents a real challenge) is easily explained by a striking feature of
the analysis adopted in Ref.~\refcite{nolose} which relies on the tracking 
devices, FPDs and muon chambers, hence reducing the effect of calorimeters to
a minimum. By selecting events with only 4 or 6 tracks and at least one muon with
$p_T > 6$~GeV, of all combinations of four tau decays, only those are chosen 
where one tau decays to a muon whilst of the three remaining taus at most one 
of them is allowed to decay to three charged particles. By insisting also that
the selected tracks have the right topology (they should cluster and form 
back-to-back pairs) we get a very favorable
ratio of signal to overlap background: while 25\% of signal is still available 
after the muon $p_T$ cut, the overlap background is immensely suppressed. 
 
The favorable signal-to-background ratio, and a signal event
yield that is not extremely low, promises good prospects not only for measuring the mass of the scalar 
Higgs boson but also of the pseudoscalar $a$. Knowing the rapidity and mass
of the central system from precise $\xi$-measurements at the FPDs, and assuming that the
$a$'s are highly boosted (which would cause their decay products to follow 
roughly the direction of the parent $\tau$'s), four $a$ mass measurements per
event are possible. With the efficiency of the muon trigger estimated to be 
around 50\% and with the amount of data of 180 \ifb collected at 
$3\,\times$~\low, one can expect roughly 6 signal events with a
negligible background, giving $m_a = 9.3\pm2.3$~GeV which is in a good 
agreement with the expected value of 9.7~GeV.

\section{\bf Central Exclusive Higgs production in Triplet Model}

In SUSY models there are at least two Higgs doublets. Singlets occur in many
extensions of the SM, see the previous section. In left-right symmetric models,
triplets are added to generate a small mass for the neutrinos. The new scalars
do not always participate in the electroweak symmetry breaking, nevertheless 
they affect the properties of the Higgs boson via mixing. Higgs triplets are
used in models such as composite Higgs or little Higgs. A tiny neutrino mass 
would indicate that the mass is generated by the seesaw mechanism which 
contains the couplings of neutrinos to the triplet. Ref.~\refcite{triplet} shows
that the lightest neutral Higgs boson of a model with triplets, $H_1^0$,  may be
favorably searched for in CEP with proposed FPDs and at the same time, it 
identifies the representation of the found $H_1^0$. 

Since the studied Higgs boson decay is the $\bb$, the same analysis 
procedure applies as for the SM or MSSM CEP of $\Hbb$ process
(see Ref.~\refcite{CLP}). The only difference is in the cross section of the
signal which in the triplet model is greatly enhanced with respect to SM:
113.5, 18.0 and 6.6~fb at $m_H = 120$~GeV for $c_H =$ 0.2, 0.5 and 0.8, 
respectively. The $c_H$ is a parameter of the triplet model specifying the 
amount of the doublet-triplet mixing. At tree level, the coupling of the 
$H_1^0$ to fermions is always enhanced by a factor of 1/$c_H$, while the gauge
boson couplings to $H_1^0$ are suppressed by a factor of $c_H$ with respect to 
the SM. Consequently the role of vector boson fusion for $H_1^0$ 
production is reduced if $c_H$ is small. It was shown in Ref.~\refcite{triplet} 
that for $c_H < 0.5$ the light Higgs boson $H_1^0$ (of mass between 120 and 
150 ~GeV) can be observed with a significance of $4\,\si$ or better and its 
mass measurement can be made of better than 2~GeV resolution, if a fixed rate 
single jet trigger is used to retain events in which both protons are measured 
at 420~m from the IP.

\section{\bf Conclusions}

Despite the achievements in terms of the discovery of the Higgs boson at LHC 
and measurements/estimates of its properties, the central exclusive production 
of the Higgs boson, whether of the SM or MSSM nature, still represents a 
powerful tool to complement the standard strategies at LHC. A striking feature 
of the CEP Higgs-boson is that this channel 
provides valuable additional information on the spin and the coupling structure of 
Higgs candidates at the LHC. We emphasize that the $J_z = 0$, $\cC$-even, 
$\cP$-even selection rule of the CEP process enables us to estimate 
very precisely (and event-by-event) the quantum numbers of any resonance 
produced via CEP. 

In this review, the current status of the signal selection and of the 
background suppression for the CEP of Higgs bosons at LHC has been summarised,
and numerous results have been collected and compared, where possible. The SM
situation has been elaborated with a special attention since it represents
a basis for all other
analyses dealing with the New Physics to be searched for using the forward
proton measurements. The significances for the CEP Higgs boson decaying into
$\bb$, $WW$ or $\tau\tau$ pairs in SM are moderate but $3\,\si$ can be reached 
if the analysis tools, ToF measurement resolution or L1 trigger strategies are 
improved. The fact that the 
mass 125.5 GeV Higgs boson is well known helps to optimize the strategy. Measurements
and estimates made at LHC so far suggest the discovered Higgs boson is very
consistent with the SM, hence room for New Physics at low masses is diminishing,
while the high mass region still needs to be explored. In this
review three models of New Physics have been examined, namely the MSSM, NMSSM 
and Triplet models. The prospects of using central exclusive production to study
this sector have been summarized, although upcoming results expected from
the LHC Run II may change the outlook.

\section*{\bf Acknowledgments}

The work was supported by the project LG13009 of the Ministry of Education of 
the Czech republic.
The author wishes to thank Valery Khoze for useful discussions, encouragement 
and assistance.


\end{document}